\documentclass[a4paper]{jpconf}
\usepackage{graphicx}
\usepackage{amssymb}
\usepackage{wrapfig}
\begin{document}
\title{MADMAX: A new road to axion dark matter detection}

\author{B. Majorovits for the MADMAX interest group$^{\small \dagger}$}

\address{Max-Planck-Institute for Physics, Werner-Heisenberg-Institute, Foehringer Ring 6, 80805 Munich, Germany}
\ead{bela.majorovits@mpp.mpg.de}
\vspace{0.2cm}
\address{{\bf $^{\dagger}$ The MADMAX interest group:} \\{\bf MPI for Radioastronomy} - Bonn, Germany: M.\,Kramer, G.\,Wieching; \\{\bf DESY Hamburg} - Hamburg, Germany: H.\,Kr\"uger, A.\,Lindner, C.\,Martens, J.\,Schaffran; \\{\bf University of Hamburg} - Hamburg, Germany: E.\,Garutti,  A.\,Schmidt; \\{\bf MPI for Physics} - Munich, Germany: A. Caldwell, G.\,Dvali, C.\,Gooch, A.\,Hambarzumjan, S.\,Knirck, B.\,Majorovits, A.\,Millar, G.\,Raffelt, O.\,Reimann, F.\,Steffen; \\{\bf CEA IRFU} - Saclay, France: P. Brun, L. Chevalier;  \\{\bf University of Zaragoza} - Zaragoza, Spain: J. Redondo}

\begin{abstract}
The axion is a hypothetical low-mass boson predicted by the Peccei-Quinn mechanism solving the strong CP problem. It is naturally also a cold dark matter candidate if its mass is below $\sim$\,1\,meV, thus simultaneously solving two major problems of nature. All existing experimental efforts to detect QCD axions focus on a range of axion masses below $\sim$\,25$\,\mu$eV. The mass range above $\sim$\,40$\,\mu$eV, predicted by modern models in which the Peccei-Quinn symmetry was restored after inflation, could not be explored so far. 
The MADMAX project is designed to be sensitive for axions with masses (40--400$)\,\mu$eV. The experimental design is based on the idea of enhanced axion-photon conversion in a system with several layers with alternating dielectric constants.
The concept and the proposed design of the MADMAX experiment are discussed. Measurements taken with a prototype test setup are discussed. The prospects for reaching sensitivity enough to cover the parameter space predicted for QCD dark matter axions with mass in the range around 100\,$\mu$eV is presented.

\end{abstract}

\section{Introduction}
Most experiments looking for dark matter axions are sensitive in the axion mass range below $\sim$\,25\,$\mu$eV \cite{exp_review}. This mass range is plausible in  
the scenario where Peccei-Quinn (PQ) symmetry was broken before inflation. So far no experimental effort has reached the sensitivity to potentially detect dark matter axions in the mass ranges 
preferred by post inflationary PQ breaking.
\cite{axion_masses, smash}.

The MADMAX experiment makes use of the dielectric haloscope concept of placing several discs made of material with high dielectric constant and low dielectric loss in front of a mirror inside a strong magnetic field \cite{dielectric, madmax}.   
The axion-photon conversion rate is proportional to the square of the magnetic field strength.
It has been argued that using 80 discs made from LaAlO$_3$ with dielectric constant $\epsilon\approx$\,24 can lead to a power boost of the signal with respect to the power emitted from a mirror by up to 5 orders of magnitudes \cite{madmax}. 
If such a power boost can indeed be achieved, a sensitivity could be reached that allows for scanning the dark matter axion mass range between $\sim$\,40\,$\mu$eV and $\sim$\,400\,$\mu$eV.

\section{Experimental Concept}
\begin{wrapfigure}{l}{0.66\textwidth}
\includegraphics[width=10cm]{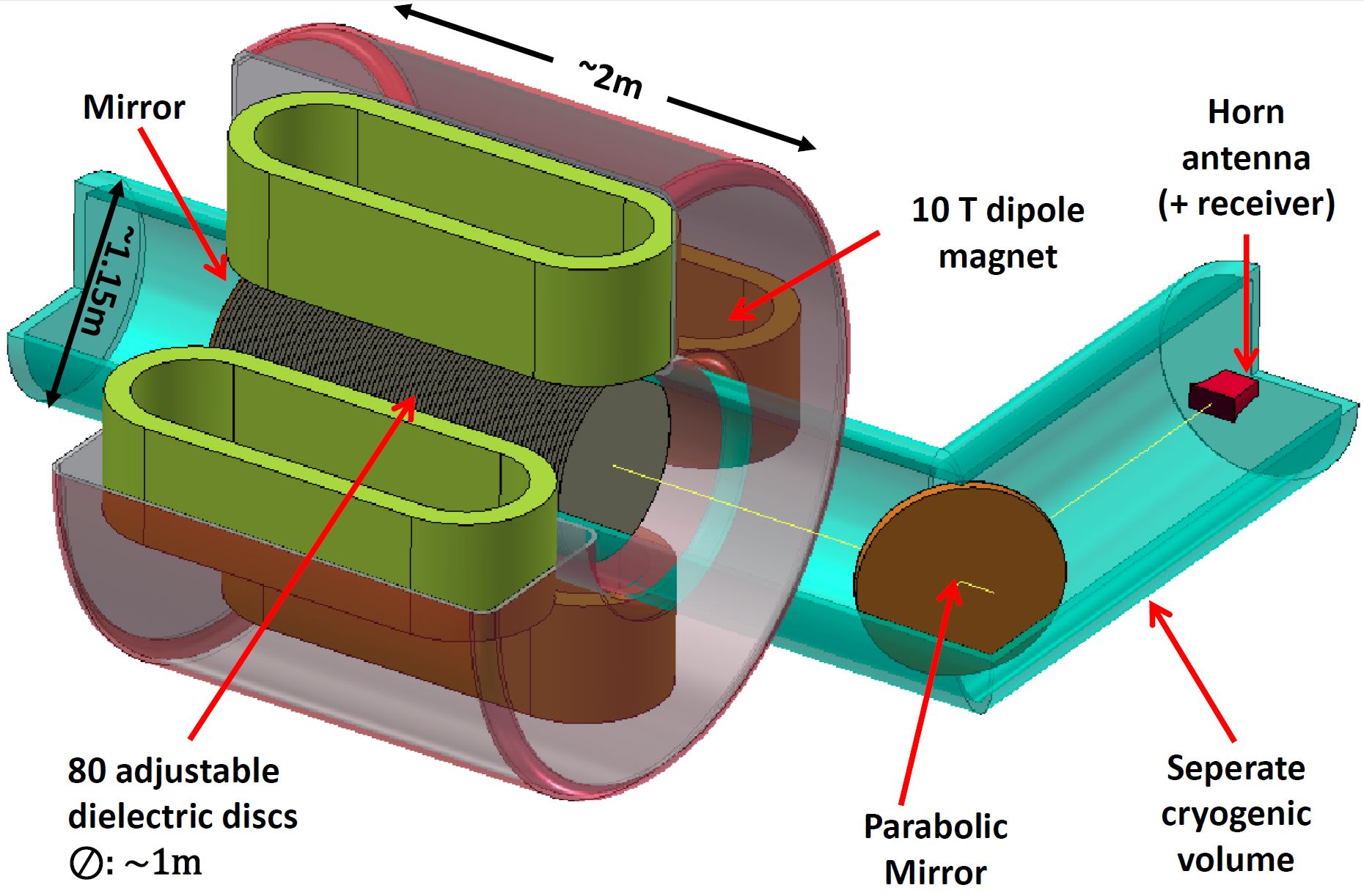}\hspace{2pc}%
\caption{\label{fig:exp}Schematic sketch of the  concept. 
}
\end{wrapfigure}

The basic concept of the experimental approach of the MADMAX experiment is depicted in Fig.\,\ref{fig:exp}. The booster consisting of one mirror and $\sim$ 80 adjustable discs is placed inside a dipole magnet with $\sim$\,10\,T field strength parallel to the surfaces of the discs. By adjustment of the distances between the inividual discs, the frequency range within which a significant power boost is obtained can be varied between 10\,GHz and 100\,GHz, corresponding to the frequencies of photons induced by $\sim$\,40\,$\mu$eV and $\sim$\,400\,$\mu$eV axions, respectively. The obtainable mean boost factor in a given frequency range is roughly inversly proportional to the bandwidth within which the boost factor is adjusted and thus enhanced \cite{foundations}. This enables a search strategy in which first a broader (tens of MHz) frequency range is investigated using a specific disc setting and all signal candidates with significance higher than a defined threshold are subsequently re-measured with an increased boost factor using disc spacings resulting in maximal resonance of the boost factor curve, i.e. minimal bandwidth and thus maximal obtainable boost factor. 
Depending on the frequency range being investigated, the disc spacings vary and the length of the booster ranges between $\sim$\,20\,cm and 2\,m. This also imposes the requirement on the length of the magnet.
The booster will be housed in a separate cryogenic volume. This allows for control of the temperature of discs, mirror, waveguide walls and mechanical infrastructure to avoid background from black body photons.
The signal emitted from the booster setup is reflected by a parabolic mirror, concentrating the power into a corrugated horn antenna at the end of the cryogenic volume. The signal can be detected using a receiver system based on heterodyne mixing of the frequency into the MHz regime.
The power boost of the booster system depends roughly linearly on the dielectric constant of the disc material \cite{foundations, QFT}. In consequence  a material with high dielectric constant $\epsilon$ should be chosen. Additionally the dielectric loss tan\,$\delta$ within the material should be as low as possible. From this point of view LaAlO$_3$ with $\epsilon$\,=\,24 and tan\,$\delta\sim10^{-5}$
at the relevant frequencies seems to be a viable choice that still needs experimental verification.
By adjusting the disc positions the frequency range within which the power boost factor is high can be adjusted between \,10\,GHz and $\sim$\,100\,GHz.

\section{First proof of principle measurement}
The predicted power of microwaves produced by dark matter axions of a disc with 1\,m$^2$ area inside a 10\,T magnetic dipole field is roughly 2\,$\cdot\,$10$^{-27}$\,W \cite{dish_antenna}. With a power boost of 10$^4$ the expected power is of the order 10$^{-23}$\,W, detectable by state of the art receivers.

\begin{figure}[h]
\includegraphics[width=8cm]{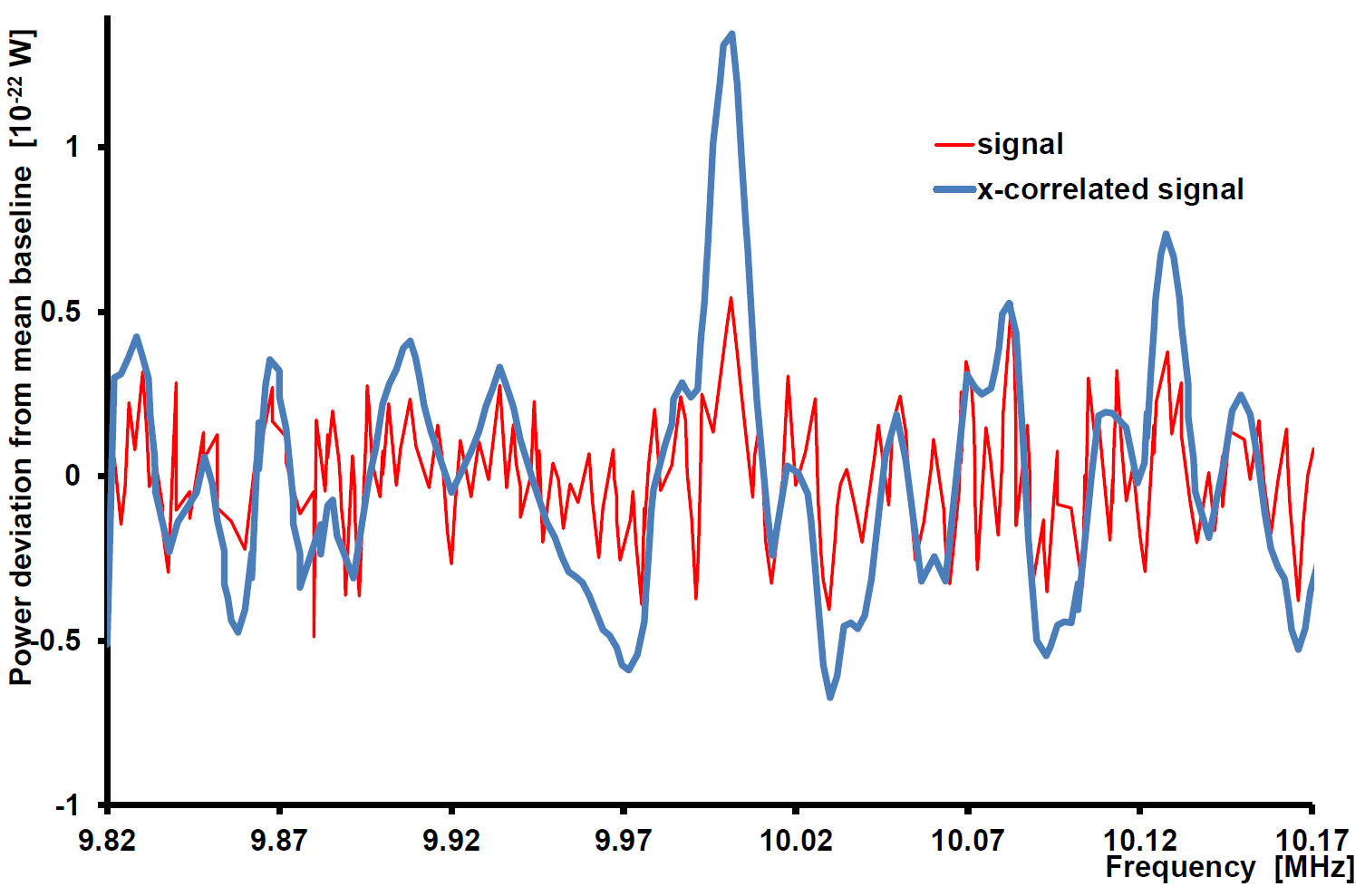}\hspace{2pc}%
\includegraphics[width=7.5cm]{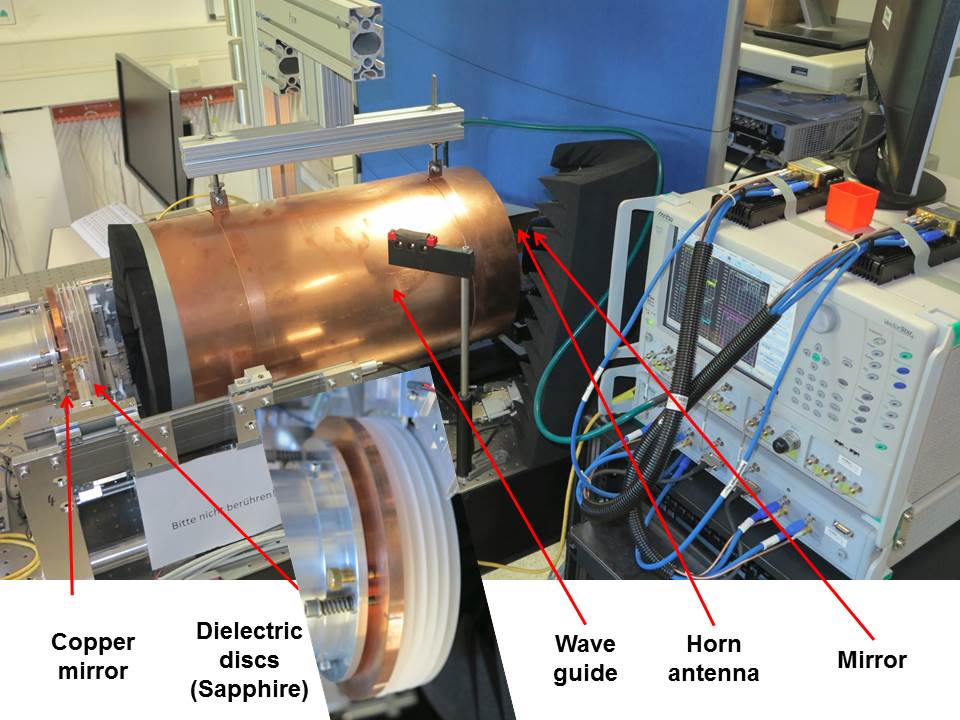}\hspace{2pc}%
\caption{\label{fig:signal}Left: Zoom into the intermediate frequency range where the 10$^{-22}$\,W signal has been identified. The red curve gives the baseline subtracted signal, while the blue curve displays the Lorentzian cross correlated signal. Right: Test setup with a mirror and five sapphire discs.}
\end{figure}

The principle of signal detection proposed for MADMAX is based on heterodyne mixing of a preamplified signal. 
After a first stage low noise preamplifier, the signal is shifted to intermediate frequencies and further amplified and filtered. A digital sampler with internal Field Programmable Gate Arrays providing real time fast Fourier transform is used to integrate and store the signal. Such a system with a HEMT as first stage preamplifier and three intermediate frequencies has been built. Depending on the HEMT preamplifier used, such a system can be sensitive in the frequency range up to $\sim$\,40\,GHz.
For higher frequencies an alternative preamplification technology still would have to be developed.
A weak signal with 17.8\,GHz and a power of 10$^{-22}$\,W has been injected into the liquid helium cooled HEMT preamplifier. After an integration time of 28 hours the signal could be detected with $>$\,6\,$\sigma$ significance using a Lorentzian cross correlation function
with a half width of 
8\,kHz. Fig.\,\ref{fig:signal}\,(left) shows the baseline subtracted signal and the cross correlated signal in the interesting frequency range. A clear peak can be observerd at the intermediate frequency corresponding to the 17.8\,GHz signal. 

\begin{figure}[t!]
\includegraphics[width=7cm]{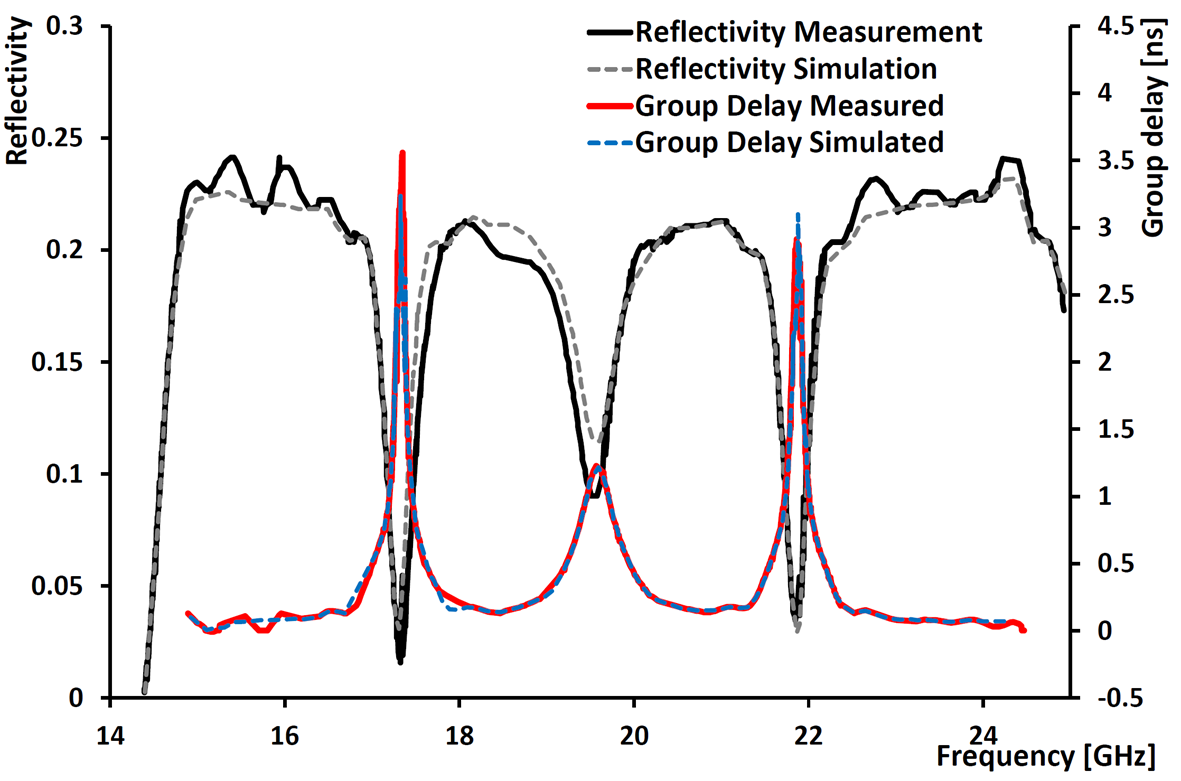}\hspace{2pc}%
\includegraphics[width=7cm]{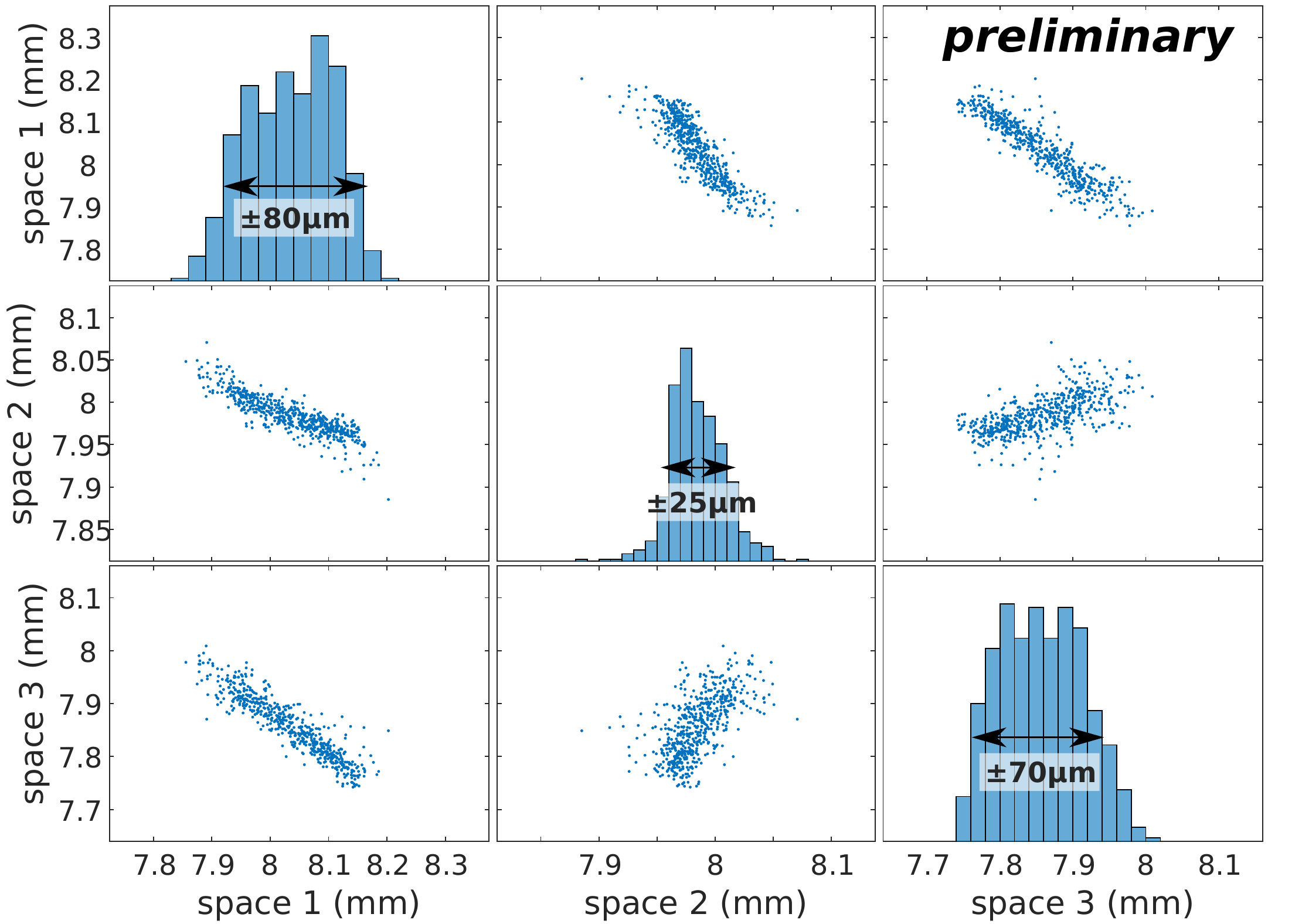}\hspace{2pc}%
\caption{\label{fig:correlations}Left: Measured and simulated group delay and reflectivity of a 3-disc-plus-mirror system. Right: Reproducibility and correlations of disc positions. The off diagonal scatter plots show the correlations between the three different disc positions after convergence of the disc placing algorithm. The diagonal histograms show the distribution of disc positions found.}
\end{figure}

In order to verify the predicted power boost behavior a smaller booster test setup with up to five sapphire discs ($\epsilon \sim$ 9, tan\,$\delta \sim 10^{-5}$ )
in front of a mirror was built. The idea is that if reflectivity, transmissivity and group delay behavior of a system are shown to agree between simulation and measurement, the boost
factor behavior of the system can be reliably obtained from classical electro-magnetic calculations as well \cite{foundations}.
The used setup is shown in Fig.\,\ref{fig:signal}\,(right). A signal with adjustable frequency can be injected into the system through a horn antenna. The signal is reflected by a parabolic mirror into the system. The response signal of the system is back-reflected by the parabolic mirror into the horn antenna and is recorded by a network analyzer. The measured group delay and reflectivity behavior as a function of frequency can be compared to simulations. This is shown in Fig.\,\ref{fig:correlations}\,(left).
The agreement between measured and simulated reflectivity and group delay curves is remarkable.

Placement of the discs happens via precision motors attached to the individual discs. In order to 
set the disc positions to achieve a desired boost factor curve,
the following algorithm has been implemented. 
In a first step the desired boost factor and frequency range are defined and 
ideal disc positions as well as the group delay and reflectivity
behavior of the system are calculated.
The discs are driven to the calculated disc position. Note
that this position is, due to mechanical tolerances, not necessarily as
precise as needed.
A reflectivity measurement is done from which the group delay is calculated. 
The group delay is compared with simulation.
If the result is improved, disc positions are saved as the last best guess.
Based on the last best guess, the disc spacings are varied randomly.
Steps are repeated iteratively. The random variations are reduced, 
if no improvement is found for a certain number of trials, until disc positions converge.

For a one-disc-plus-mirror setup the reproducibilty of the disc position of about
2\,$\mu$m could be achieved.
For a three-disc-plus-mirror system
the reproducibility and correlations of positions of individual discs are   
shown in Fig.\,\ref{fig:correlations}\,(right).
It is clearly visible that the solutions found are highly correlated among the discs.
While the reproducibilty of the indvidual positions is of the order tens of $\mu$m,
nevertheless the group delay as a function of frequency and hence the boost factor curves could be
reproduced reliably. Work is presently in progress to understand the
behavior of systems with an increased number of discs.

\begin{wrapfigure}{r}{0.5\textwidth}
\hspace{0.5cm}
\includegraphics[width=7cm]{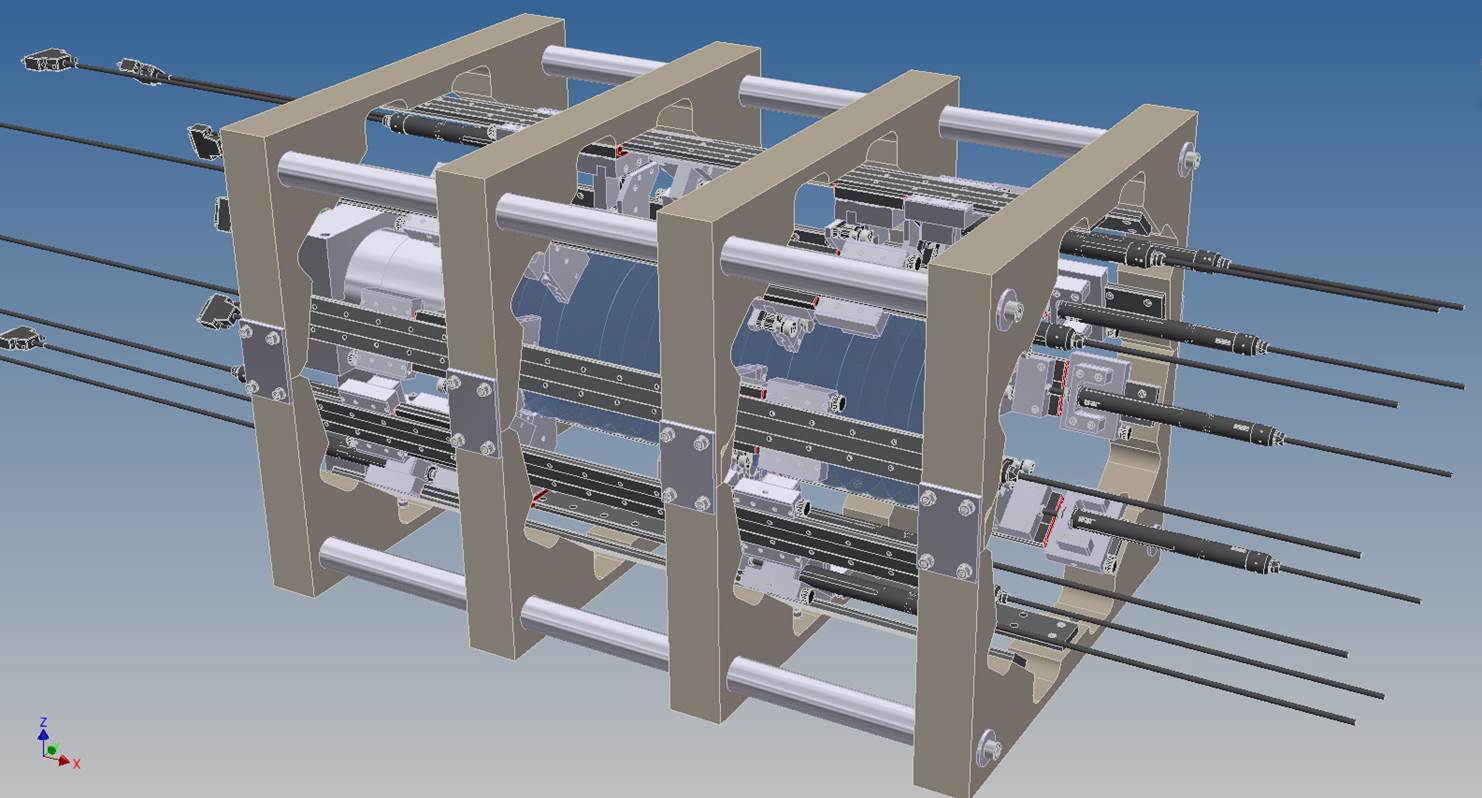}\hspace{2pc}%
\caption{\label{fig:setup}Sketch of the test setup for up to 20 sapphire discs.}
\end{wrapfigure}

Fig.\,\ref{fig:setup} shows the sketch of an upgraded setup that allows for investigation up to 20
discs. This setup is presently under commissioning at the MPP.

\section{Outlook and expected sensitivity}
Presently the viability of building a dipole magnet with required field strength and aperture is being investigated in two independent design studies. A realistic implementation could be based on a racetrack design with an active shield \cite{racetrack}. As a first step it is planned to build a prototype magnet with smaller aperture and a magnetic dipole field of $\sim$\,3\,T.

Before building a full scale experiment, it is planned to test the mechanical feasibility using a prototype booster with $\sim$ 20 discs of diameters $\sim$ 30\,cm inside the prototype magnet.
Also it needs to be understood how the physical temperature of the setup influences the noise temperature of the receiver system.
This will be done by performing noise measurements of different waveguides and disc setups inside a 
dedicated liquid helium cryostat.

\begin{wrapfigure}{r}{0.65\textwidth}
\includegraphics[width=10.5cm]{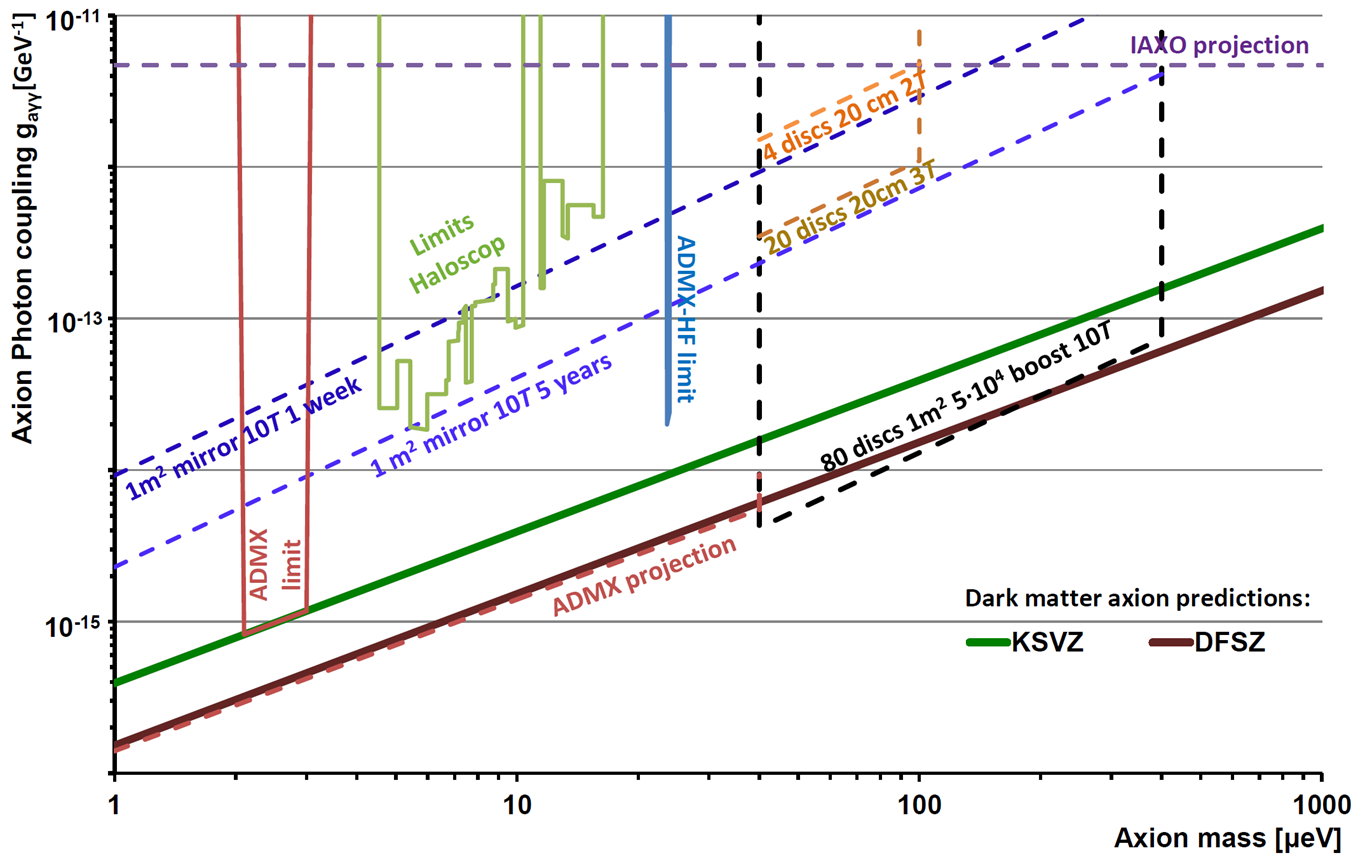}\hspace{2pc}%
\caption{\label{fig:sensitivity} Sensitivity estimate of the MADMAX approach in the axion-photon coupling vs. axion mass range compared to existing limits and senstivities of other approaches. Details can be found in \cite{white_paper}.}
\end{wrapfigure}

Once the concept has been verified for a larger amount of discs and a larger disc diameter, the full
scale experiment could be built. DESY Hamburg is presently discussed as a possible site.
The expected sensitivity of a system with 80 discs with 1\,m$^2$  area each inside a 10\,T field is shown in Fig.\,\ref{fig:sensitivity}. Here it is assumed that a power boost of 5\,$\cdot\,10^4$ can be achieved for a bandwidth of 40\,MHz and a frequency stability of the boost factor behavior of 1\,MHz. 
The signal loss due to diffraction is assumed to be negligible. A noise temperature of 8\,K and a signal width of 10$^{-6}\nu$ is assumed.
With above assumptions, using Dickes radiometer equation \cite{dicke}, in each 40\,MHz window within 7\,days of measuring time a senstivity could be reached that allows for scanning the KSVZ and DFSZ model specific predictions for dark matter axions.
Fig.\,\ref{fig:sensitivity} also shows existing limits and projected sensitivities of other proposals or technologies.
Assuming conservatively that the time needed to re-arrange the disc settings of the booster for different frequency ranges
or for improving the boost factor by narrowing the frequency range equals one day, a total time of $\sim$\,80 days will be needed
to scan 1\,GHz of bandwidth and re-scan possible candidates with $\gtrsim$4$\sigma$ significance. In consequence it would take approximately 4 years to scan the frequency range between 10 and 30\,GHz, corresponding to the mass range between 40 and 120\,$\mu$eV.

\section{Conclusions}
Presently there are no dark matter axion experiments that are sensitive to the mass range predicted by the scenario where Peccei-Quinn symmetry breaking happened after inflation. The MADMAX experiment could fill this gap by exploiting the dielectric haloscope method. First proof of principle calculations and comparison to measurements with a down sized setup 
containing five sapphire discs with 20\,cm diameter in front of a mirror are encouraging. Extrapolating the obtained results to a larger setup with 80 discs with 1\,m$^2$ surface in a 10\,T magnetic field results in an experiment that could be sensitive to the parameter range predicted by most common dark-matter axion models.

\section{Acknowledgments}
This work was partially supported the Excellence Cluster Universe.


\section*{References}
\begin{thebibliography}{9}
\bibitem{exp_review} Graham P et al.,  2015 {\it Annu. Rev. Nucl. Part. Sci} {\bf 65}: 485-514  
\bibitem{axion_masses} Borsanyi S et al., 2016 {\it Nature} \textbf{539} 69
\bibitem{smash}  Ballesteros G et al., 2017 {\it JCAP} {\bf 08} 001 [arXiv:1610.01639]
\bibitem{dielectric} Jaeckel J and Redondo J, {\it Phys. Rev.} D 2013 \textbf{88} 115002 [arXiv:1308.1103]
\bibitem{madmax} The MADMAX working group,  Caldwell A et al., 2017 \PRL {\bf 118} 091801 , [arXiv:1611.05865]
\bibitem{foundations} Millar A et al., 2017 {\it JCAP} {\bf 1701} 061, [arXiv:1612.07057]
\bibitem{QFT} Ioannisian A et al., 2017 {\it JCAP} {\bf 09} 005, [arXiv:1707.00701]
\bibitem{dish_antenna} Horns D et al., 
2013 {\it JCAP} \textbf{1304} 016 [arXiv:1212.2970].  
\bibitem{racetrack}  Rochepault E, Vedrine P and Bouillault F, 2012 {\it IEEE Trans.  Appl. Supercond} \textbf{22} 4900804
\bibitem{dicke} Hunter T and  Kimberk R, Procs. of  26th Int. Symp. on Space THz Tech., 
[ arXiv:1507.04280]
\bibitem{white_paper} The MADMAX interest group, Caldwell A et al., https://www.mpp.mpg.de/fileadmin/user\_upload/
Forschung/MADMAX/madmax\_white\_paper.pdf
\end{thebibliography}
\end{document}